# Strain-programmable exciton diffusion in moiré heterostructures

Chiho Song,[1,†] Chiranjit Mondal,[1,2,†] Jaebin Lee,[1] Kenji Watanabe,[3] Takashi Taniguchi,[4] Bohm-Jung Yang,[1,2,5] and Jieun Lee[1,5,*]
[1]Department of Physics and Astronomy, Seoul National University, Seoul 08826, Korea.
[2]Center for Theoretical Physics, Seoul National University, Seoul 08826, Korea.
[3]Research Center for Electronic and Optical Materials, National Institute for Materials Science, Tsukuba 305-0044, Japan.
[4]Research Center for Materials Nanoarchitectonics, National Institute for Materials Science, Tsukuba 305-0044, Japan.
[5]Institute of Applied Physics, Seoul National University, Seoul 08826, Korea.

[†]These authors contributed equally to this work.
*Corresponding author: lee.jieun@snu.ac.kr

Moiré superlattices in van der Waals heterostructures have recently gained significant attention as an intriguing platform for studying correlated electronic systems and exotic excitonic properties. Previous reports, however, focused on creating and modulating moiré heterostructures through interlayer twisting or lattice constant mismatches, limiting controls on symmetry of heterostructures. In this work, we show that strain significantly alters the geometry of moiré superlattices by breaking the $C_3$ rotational symmetry. We realize strain-induced moiré superlattices by intentionally regulating interlayer strain in $WSe_2$-$MoSe_2$ heterostructures, which is manifested by linearly polarized interlayer exciton emission coupled to the strain direction. Furthermore, interlayer exciton diffusion was preferentially guided along the stretched moiré superlattice orientations over a wide spatial range, reflecting the strain-modified moiré potentials. Our work highlights strain tuning as a versatile tool for designing moiré superlattices and programming excitonic transport, which opens pathways for van der Waals logic and information processing devices.

van der Waals (vdW) heterostructures based on two-dimensional (2D) materials have recently emerged as a fascinating system for quantum phenomena and correlated electron physics leading to various novel phases of matter [1,2]. For example, vdW moiré heterostructures host highly tunable correlated excitonic states with great potential for quantum simulators and advanced nano-optoelectronics [3-6] . The moiré periodicity ($\lambda$) in previous works was generally controlled by changing the twist angle ($\theta$) or introducing lattice constant mismatch ($\delta$) between layers through the relation $\lambda = a/\sqrt{\theta^2+\delta^2}$, where $a$ is the lattice constant. Such tuning has enabled a wide range of emerging phenomena in vdW heterostructures, but without changing the symmetry of the moiré superlattices [7-9].

Distinct from previously studied twisted systems, strain also offers a powerful tool for tuning moiré superlattices in vdW heterostructures. Theory showed that diverse moiré patterns including square, rectangular, or even one-dimensional (1D) moiré stripes can be created by strain [10,11], enabling a wide modulation of band structures and anisotropic correlations [11-15]. For example, unconventional many-body systems can be realized such as 1D Luttinger chains [16]. Strained moiré systems are anticipated to provide versatile platforms for examining designer quantum materials to explore correlated and topological phenomena in arbitrarily programmed superlattices [17-19]. However, due to the difficulty of controlling strain on a large scale and the effect of nonuniform strain fields frequently arising during the fabrication process [20], strain-induced generation of long-range moiré superlattices has not yet been experimentally realized. Several works have recently investigated strategies to introduce and control strain in 2D heterostructures through substrate bending [21], tip-induced mechanical control [22], and epitaxial growth engineering [23], but long-range strain-induced moiré heterostructures and their strain-dependent phenomena remain experimentally unexplored.

In this work, we explore the strain-induced moiré superlattices in $WSe_2$-$MoSe_2$ heterostructures through the combination of optical measurements, scanning probe microscopy, and theoretical simulations. By introducing differential strain between two constituent monolayers, we experimentally realized moiré superlattices with anisotropic periodic modulations.

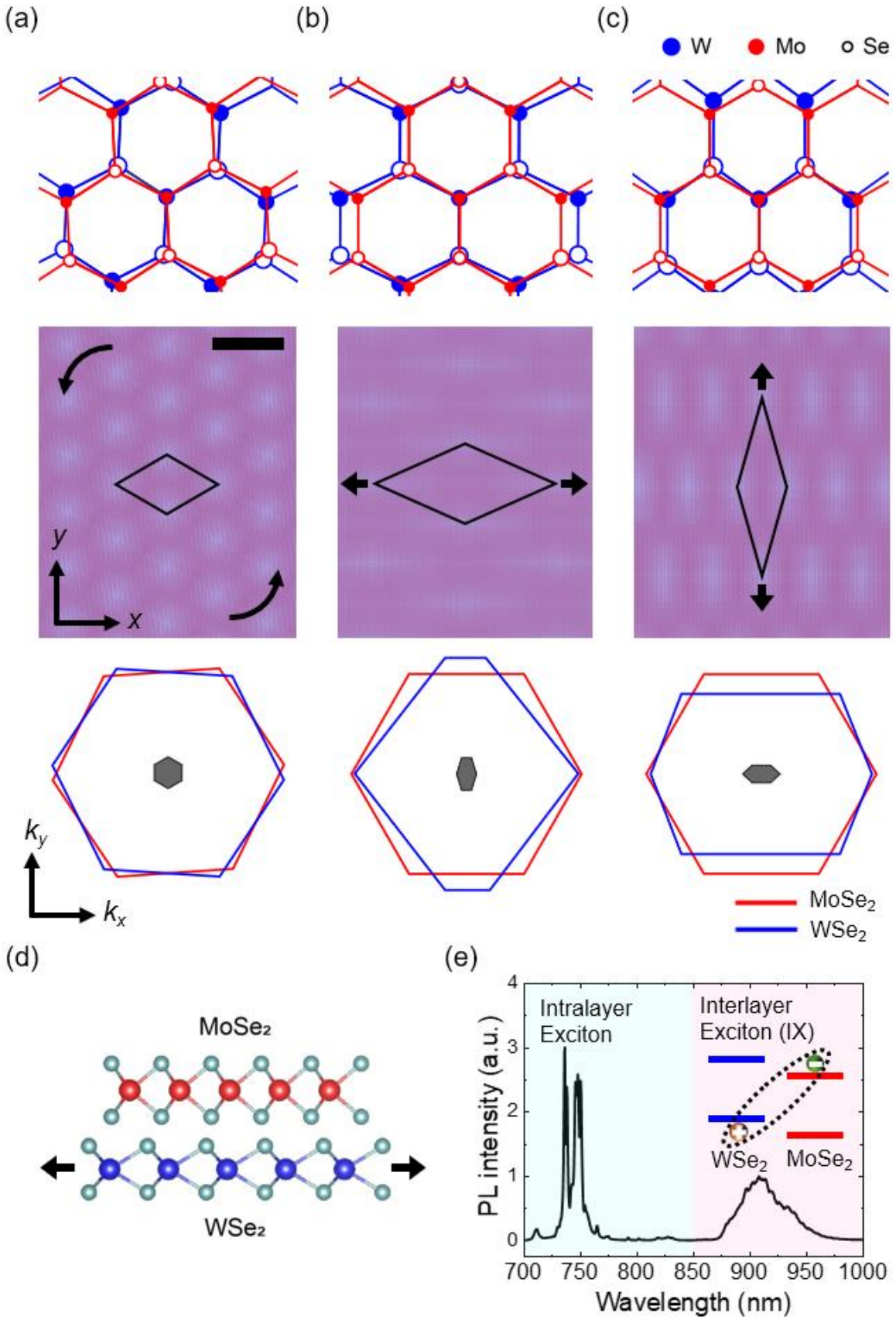

FIG. 1. Strain-induced moiré superlattices. (a–c) Upper: Stacked crystal structures of twisted (a), ZZ-strained (b), and AC-strained (c) $WSe_2$-$MoSe_2$ heterostructures. Middle: Corresponding moiré superlattices with unit cells indicated as a guide to eyes. Scale bar is 200 nm. Lower: Monolayer $MoSe_2$ (red) and $WSe_2$ (blue) BZs plotted together with the moiré BZs (gray) in reciprocal space. (d) Schematic of $WSe_2$-$MoSe_2$ heterostructure with uniaxial strain applied to bottom $WSe_2$ layer. (e) PL spectrum of intra- and interlayer excitons in sample A. Inset: Type-II band alignment of $WSe_2$-$MoSe_2$ heterostructure and interlayer exciton (IX) formation.

The spatial uniformity of strained moiré heterostructure is verified by linearly polarized interlayer excitons (IXs) coupled to the applied strain directions over a wide spatial range. Spatial imaging of IX diffusion further demonstrates IX transport that is selectively guided towards the applied strain direction, compatible with the diffusion simulations. Our results demonstrate strain control for manipulating moiré superlattice geometries, opening up opportunities for tailoring excitonic properties [24] and reconfiguring many-body interactions [3] in vdW heterostructures.

The crystal structures of strain-induced moiré heterointerfaces by applying differential interlayer strain are illustrated in Fig. 1(a–c). Unlike twisted moiré heterostructures, strain-induced moiré structures have reduced symmetry due to highly anisotropic lattice constant mismatches. Because of the small lattice constant difference between $MoSe_2$ and $WSe_2$, applying uniaxial strain of even 0.2 % to bottom $WSe_2$ already dramatically modifies moiré heterostructures. By choosing the strain direction along either zigzag (ZZ) or armchair (AC), the spatial elongation of superlattice becomes markedly contrasting. Also, the strain-induced moiré Brillouin zones (BZs) illustrated by the gray shaded areas at the centers of the monolayer BZs represent the modulation of the reciprocal moiré supercells with elongation directions linked to the strain directions.

In our experiments, we fabricated strain-induced moiré heterostructures by sequentially transferring pristine $MoSe_2$ onto strained $WSe_2$ with aligned crystal axes (Fig. 1(d)). The bottom $WSe_2$ exfoliated onto ozone-treated flexible substrate was uniaxially strained before transfer onto $SiO_2$/Si substrates. More details on sample fabrication can be found in Supplemental Material S1. The strain level was controlled using a micro-manipulator and monitored through photoluminescence (PL) measurements (see Supplemental Material S3). The top $MoSe_2$ layer was transferred without strain, producing a combined structure in a heterostrain configuration. The images of samples can be found in Supplemental Material S4. The precise angle alignment between the two layers was verified by polarization-resolved second harmonic generation (SHG) measurements, confirming the angle mismatch less than or equal to 1.7° (see Supplemental Material S5). The stacking order of all samples was determined to be R-configuration from the IX *g*-factor measurements (see Supplemental Material S6).

Fig. 1(e) shows the PL spectrum of a ZZ-strained $WSe_2$-$MoSe_2$ heterostructure (sample A) exhibiting IX emission around 920 nm [25]. Its strain level was 0.32 % as extracted during the fabrication process. Spatially scanned PL measurements confirmed the homogeneity of IX formation over the stacked region, spanning an area larger than 20 $\mu m^2$ (see Supplemental Material S7). The sizable IX emission intensity, comparable to that of intralayer excitons, further manifests strong interlayer tunneling strength, indicating a minimal twist angle [26,27].

We next measured linearly polarized IX emission intensity in sample A as a function of the polarization angle (Fig. 2(a)). A continuous-wave (CW) 532 nm laser with linear polarization was focused onto the sample and a half-wave plate and an analyzer in the collection path selectively measured the linearly polarized IX luminescence. From the measurement, we found that the emitted PL is linearly polarized with the maximum intensity axis perpendicular to the strain direction.

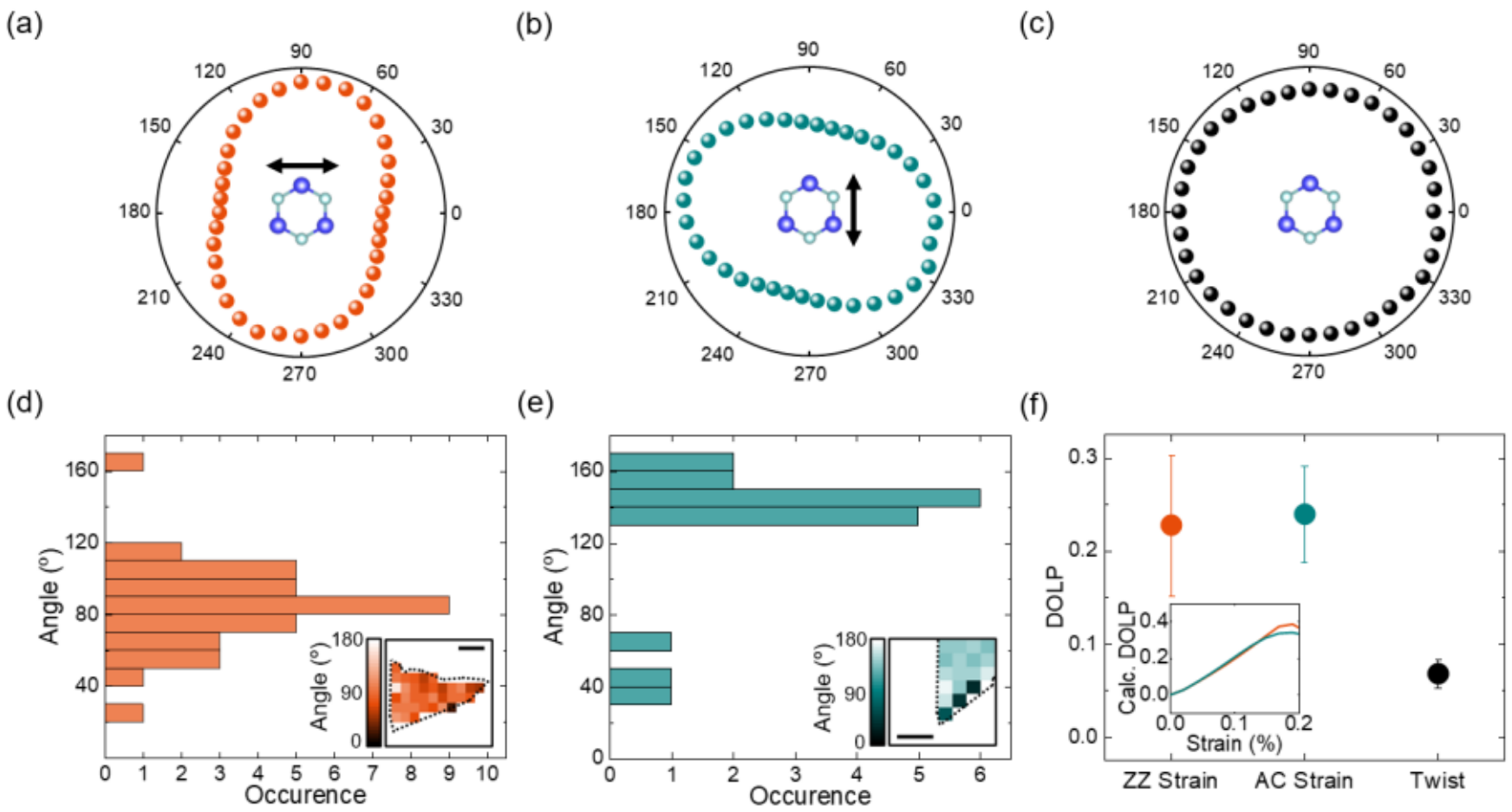

FIG. 2. Linearly polarized IX emission. (a–c) Linear polarizations of IX in ZZ-strained (a), AC-strained (b), and unstrained (c) heterostructures. Black arrows indicate the applied strain directions. (d, e) Histograms of linear polarization directions in ZZ-strained (d) and AC-strained (e) samples. Insets: Position-dependent linear polarization directions collected from the corresponding heterostructure areas. Scale bars are 5 μm in (d) and 2 μm in (e). (f) Averaged degree of linear polarization (DOLP) defined as $(I_{max} - I_{min})/(I_{max} + I_{min})$ for ZZ-strained, AC-strained, and unstrained (twisted) samples. Inset: Calculated DOLP as a function of strain level along ZZ and AC directions.

Furthermore, the PL polarization axis remained remarkably uniform over the stacked heterostructure regions, as shown by the spatial scan and histogram data plotted in Fig. 2(d). This observation suggests 1) the formation of strain-induced moiré superlattices across the sample and 2) the correlation between the linearly polarized IX emission and the applied strain direction. Piezoresponse force microscopy (PFM) on a similarly strained sample further confirms the formation of elongated moiré superlattices by our method (see Supplemental Material S8).

To validate IX linear polarization intertwined to external strain, we fabricated another sample strained along the AC-direction with a strain level of 0.14 % (sample B). Fig. 2(b) and 2(e) show that sample B also exhibits linearly polarized IX emission with the high uniformity of polarization axis across the heterostructure area. The measured linear polarization axis was also close to the perpendicular direction of the applied strain direction, consistent with the results from sample A. By contrast, such linearly polarized PL was not observed in an unstrained heterostructure (sample C), as depicted in Fig. 2(c). Comparing the degree of linear polarization (DOLP) in ZZ-strained, AC-strained, and unstrained samples (Fig. 2(f)), the strained heterostructures exhibited DOLP much higher than the unstrained one.

The emergence of the linear polarization of IXs by strain can be understood through the broken $C_3$ rotational symmetry of the moiré superlattice in reciprocal space. As illustrated in Fig. 1(e), the type-II band alignment of $WSe_2$-$MoSe_2$ heterostructure induces IXs consisting of holes and electrons residing in the valence and conduction bands of $WSe_2$ and $MoSe_2$, respectively. When the two layers are twisted, the displacement vectors between the two opposite band extrema at K and −K points with the same magnitude develop in all BZ corners. Since photons with opposite helicity couple to moiré excitons at K and −K points, the DOLP of twisted moiré superlattice is nearly zero under linearly polarized excitation laser. However, strain modifies the optical transition at BZ corners due to the broken rotational symmetry, and with the additional distortion of IX wavefunction distributions induced by strain, a finite DOLP emerges in strained heterostructures [28,29].

We calculated the DOLP for $WSe_2$-$MoSe_2$ heterostructures considering the moiré exciton wavefunctions under varying strain, which is shown in the inset of Fig. 2(f). The calculated DOLP increases with the strain level, agreeing well with our experiment, supporting that IXs are subjected to strain-induced moiré supercells. Moreover, our calculation shows that in addition to ground-state moiré excitons, excited-state moiré excitons are involved for the observation of the polarization axis perpendicular to the strain directions. Considering the laser power used (10 μW), it is highly possible that excited-state IXs are populated, in agreement with previous literature [30]. More calculational results are provided in Supplemental Material S9.

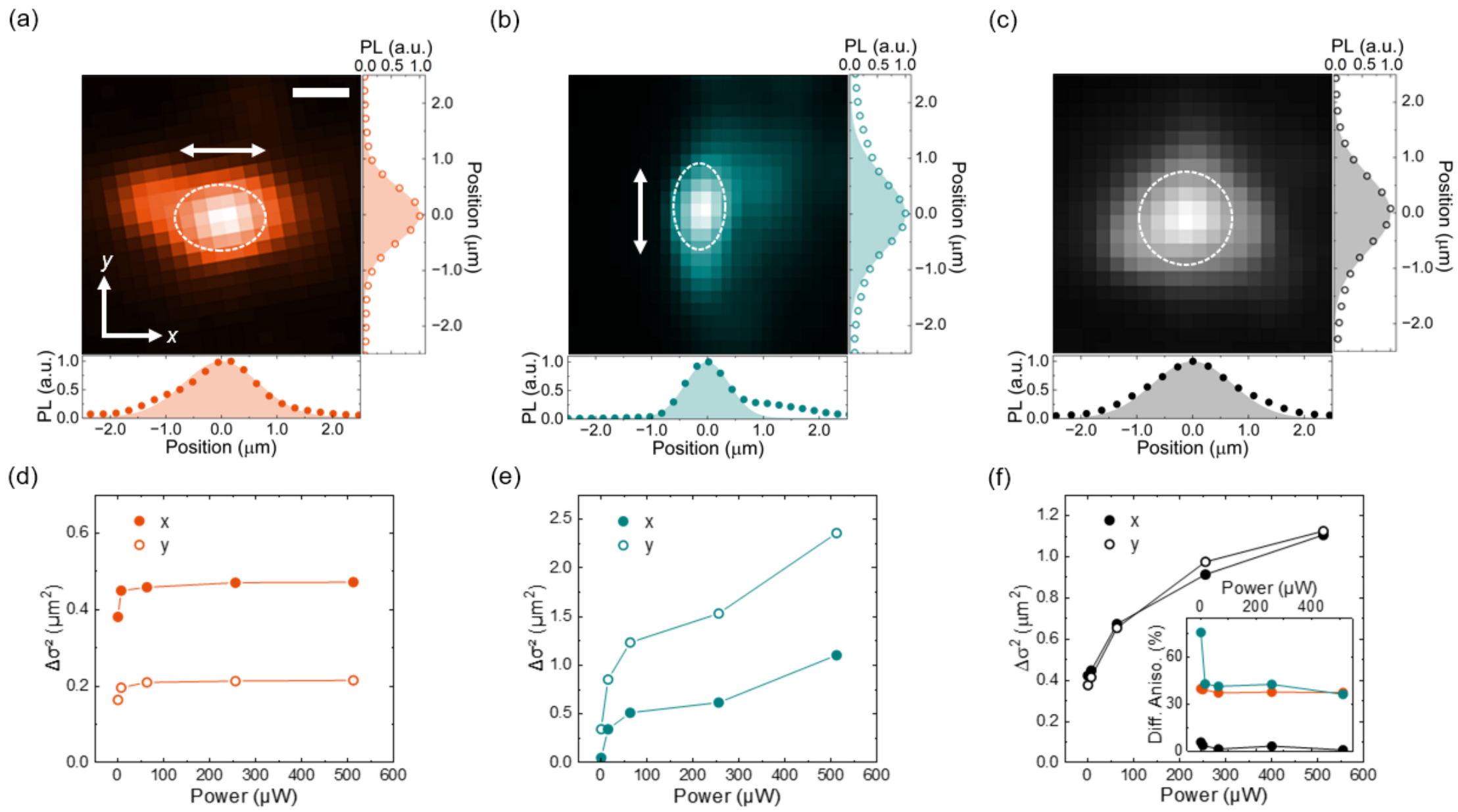

FIG. 3. Anisotropic IX diffusion. (a–c) Upper: Spatial PL maps of IX diffusion in ZZ-strained (a), AC-strained (b), and unstrained (c) heterostructures at the excitation laser power of 1 µW, with the line profiles along the $x$- and $y$-axis in corresponding heterostructures. White arrows indicate the applied strain directions. Dashed lines are a guide to eyes. Scale bar is 1 µm. Gaussian fitting functions are plotted as shaded areas. (d–f) Power dependence of the differential squared width $\Delta\sigma^2$ along $x$- and $y$-axis in ZZ-strained (d), AC-strained (e), and unstrained (f) heterostructures. Inset: Diffusion anisotropy as a function of power for ZZ-strained (orange), AC-strained (dark cyan), and unstrained (black) samples.

Having established heterostructures with uniformly strained moiré superlattices, we next investigated strain-modulated transport properties of IXs. For these measurements, the heterostructures were excited using a laser focused by an objective lens, yielding a spot size ($\sigma_0$) of about 0.3 µm, and the spectrally filtered IX emission was mapped using the imaging mode of a charge-coupled device (CCD). A PL map of sample A is shown in Fig. 3(a), revealing the IX diffusion beyond the excitation source area. Surprisingly, IX diffusion is strongly enhanced along the ZZ-orientation, parallel to the strain direction. For AC-strained sample B, the IXs are found to be preferentially transported along the AC-direction (Fig. 3(b)). However, no directional IX diffusion is detected in unstrained sample C (Fig. 3(c)).

To characterize the observed anisotropy in IX diffusion, 1D spatial profiles of PL along two perpendicular axes are plotted in Fig. 3(a–c). By fitting to a Gaussian function, we extract the steady-state diffusion length ($\sigma_{IX}$) along each axis, from which the differential squared width of IXs is obtained: $\Delta\sigma^2 = \sigma_{IX}^2 - \sigma_0^2$. Fig. 3(d) and 3(e) show the pump power-dependent evolution of $\Delta\sigma^2$ for ZZ- and AC-strained samples, respectively, representing the increased IX transport along the strain orientation at all powers. We then calculated the diffusion anisotropy, defined by $(\Delta\sigma_\parallel^2 - \Delta\sigma_\perp^2)/(\Delta\sigma_\parallel^2 + \Delta\sigma_\perp^2)$, using the diffusion lengths parallel ($\sigma_\parallel$) and perpendicular ($\sigma_\perp$) to the strain direction, as shown in the inset of Fig. 3(f). These results demonstrate the anisotropic IX transport which is substantially controlled using strain as a tuning knob.

To understand the origin of anisotropic IX diffusion, we performed numerical simulations of the exciton dynamics under strain-induced moiré potentials. In the simulations, the evolution of the IX density $n(\mathbf{r},t)$ under the moiré potential $u(\mathbf{r})$ at temperature $T$ is obtained by solving the Fokker-Planck equation [31]:

$$\frac{\partial n}{\partial t} = D_{\mathrm{IX}}\nabla^2 n + \frac{D_{IX}}{k_B T}(n\nabla^2 u + \nabla n\cdot\nabla u) - \frac{n}{\tau_{IX}} \quad (1)$$

where $D_{IX}$, $\tau_{IX}$, and $k_B$ are the isotropic IX diffusion coefficient, IX lifetime, and Boltzmann constant, respectively. The moiré potential $u(\mathbf{r})$ is obtained by the density functional theory (DFT) calculations incorporating local stacking configurations and interlayer distance variations of the heterostructure (see Supplemental Material S10). To obtain $n(\mathbf{r},t)$ and solve the boundary condition problem with moiré periodicity, we further implemented the pseudo-spectral time domain method to handle many spatial grids associated with oscillating potentials. More simulation details can be found in Supplemental Material S11.

In Fig. 4, we show the calculated time-integrated diffusion profiles of IXs under ZZ-strained, AC-

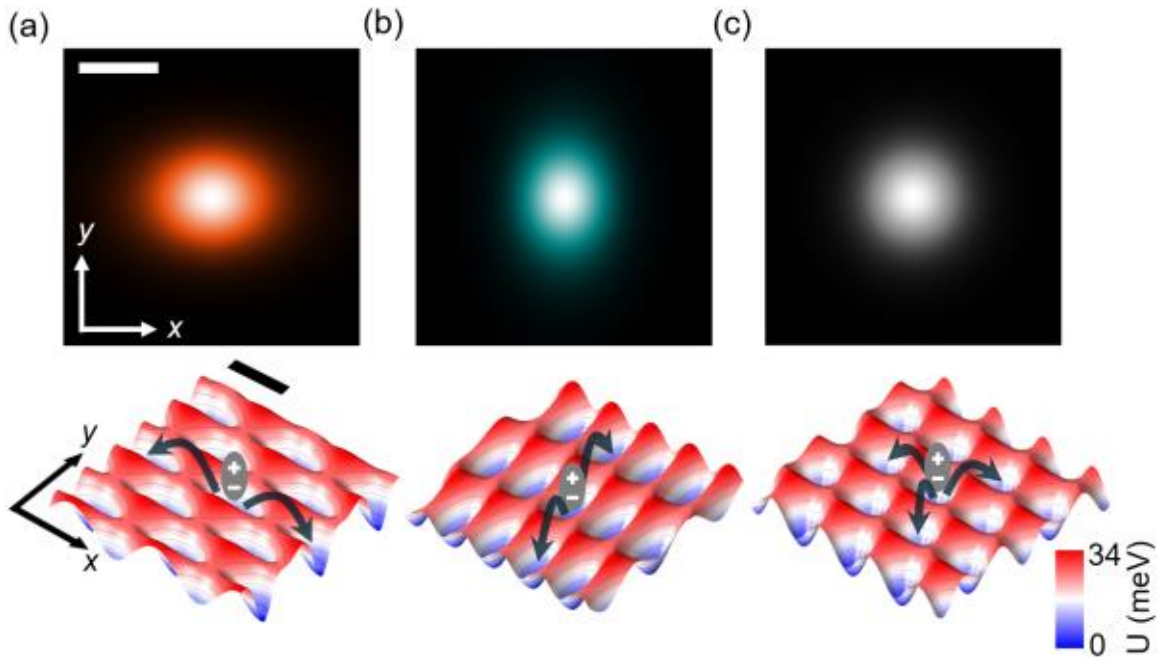

FIG. 4. Simulations of IX diffusion and moiré potential landscapes. (a–c) Upper: Calculated spatial steady-state IX distributions in ZZ-strained (a), AC-strained (b), and twisted (c) heterostructures. Scale bar is 0.5 µm. Lower: Moiré potential landscapes plotted for corresponding heterostructures. Scale bar is 50 nm. Black arrows indicate preferred IX hopping directions.

strained, and unstrained conditions. Clear anisotropy of IX diffusion is observed for both the ZZ- and AC-strained heterostructures. The diffusion anisotropy obtained from the simulation was 0.24 (ZZ) and 0.25 (AC) under a strain level of 0.3 %. The resemblance of the simulation with the experiment is striking, considering the absence of fitting parameters except for $D_{IX}$ and $\tau_{IX}$ both extracted from measurements. For the unstrained heterostructure, the simulated IX density shows an isotropic diffusion profile, consistent with the rotationally symmetric moiré potential map.

Microscopically, the strain-driven diffusion of IXs can be understood through IX transport that depends highly on the hopping strength of carriers between adjacent moiré trap sites [32]. For minimally twisted systems, a sizable moiré cell dimension (~ 20 nm) leads to deep moiré potentials (~ 100 meV) prohibiting the delocalized spreading of IX states [32]. By introducing strain, hopping-driven transport becomes more probable along the strain orientation, because the slopes of potential barriers are less steep along this direction, as visualized from the simulated moiré potential landscapes (plotted together in Fig. 4).

Our analysis on anisotropic IX diffusion under the moiré potential landscape is further supported by the power-dependent measurements of $\Delta\sigma_{\parallel}^2$ and $\Delta\sigma_{\perp}^2$. From Fig. 3(d) and 3(e), $\Delta\sigma_{\parallel}^2$ is larger than $\Delta\sigma_{\perp}^2$ for both ZZ- and AC-strained samples in all power regimes. However, the IX diffusion in sample A shows obvious saturation with increasing excitation laser power, whereas that in sample B does not. We attribute this difference to the smaller twist angle of sample A compared to sample B (see Supplemental Material S5). Since the moiré periodicity of the former (~ 23 nm) is larger than that of the latter (~ 11 nm), IXs with the radius of a few nm are more effectively confined within the moiré potential wells in sample A [30,33]. More interestingly, power-dependent measurements show the abrupt decrease of diffusion anisotropy as the laser power increases (inset of Fig. 3(f)). As exciton-exciton repulsion increases at higher laser powers [34], the strain-induced anisotropy of the IX diffusion decreases, which is more pronounced for sample B with less confinement.

In conclusion, we have investigated the generation of strain-induced moiré superlattices using layer-selective strain control of $WSe_2$-$MoSe_2$ heterostructures. We observed the linearly polarized IX emission homogeneously generated over the stacked region, demonstrating the emergence of the strained moiré superlattices over a wide spatial range. We further revealed the directional IX diffusion along the applied strain direction, which originates from the elongated moiré potential wells. The strain-induced anisotropy of IX diffusion is theoretically supported by numerical simulations of excitonic transport. With the advancement of strain control, we expect our results to be applicable for studying moiré physics in unexplored regimes, such as strain-induced topological phase transitions, vdW quantum optomechanics, and dynamically tunable moiré logic devices.

## ACKNOWLEDGMENTS

We thank Euijoon Kwon for valuable discussions on the pseudo-spectral time domain method. We also acknowledge support from the National Research Foundation of Korea (NRF) grant funded by the Korean government (MSIT) (No. 2021R1A5A103299614, No. RS-2024-00356893, and No. RS-2024-00413957). The research is further supported by the Institute for Basic Science (IBS) in Korea (No. IBS-R009_D1) and the Information Technology Research Center (ITRC) Support Program funded by the Institute for Information & Communications Technology Promotion (IITP) of Korea (No. RS-2022-00164799). C.M. and B.J.Y. acknowledge support from the Samsung Science and Technology Foundation (No. SSTF-BA2002-06), National Research Foundation of Korea (NRF) funded by the Korean government (MSIT) (No. RS-2021-NR060087), Global Research Development Center (GRDC) Cooperative Hub Program through the NRF funded by the MSIT (No. RS-2023-00258359), and Global-LAMP program of the NRF funded by the Ministry of Education (No. RS-2023-00301976).

[1] K. S. Novoselov, A. Mishchenko, A. Carvalho, and A. H. Castro Neto, Science **353**, aac9439 (2016).
[2] Y. Liu, N. O. Weiss, X. D. Duan, H. C. Cheng, Y. Huang, and X. F. Duan, Nat. Rev. Mater. **1**, 16042 (2016).

[3] D. M. Kennes, M. Claassen, L. Xian, A. Georges, A. J. Millis, J. Hone, C. R. Dean, D. N. Basov, A. N. Pasupathy, and A. Rubio, Nat. Phys. **17**, 155 (2021).
[4] L. Du, M. R. Molas, Z. Huang, G. Zhang, F. Wang, and Z. Sun, Science **379**, eadg0014 (2023).
[5] A. Ciarrocchi, F. Tagarelli, A. Avsar, and A. Kis, Nat. Rev. Mater. **7**, 449 (2022).
[6] Y. Liu, A. Elbanna, W. Gao, J. Pan, Z. Shen, and J. Teng, Adv. Mater. **34**, e2107138 (2022).
[7] X. Sun *et al.*, Chem. Rev. **124**, 1992 (2024).
[8] P. Törmä, S. Peotta, and B. A. Bernevig, Nat. Rev. Phys. **4**, 528 (2022).
[9] E. Y. Andrei, D. K. Efetov, P. Jarillo-Herrero, A. H. MacDonald, K. F. Mak, T. Senthil, E. Tutuc, A. Yazdani, and A. F. Young, Nat. Rev. Mater. **6**, 201 (2021).
[10] Q. Tong, H. Yu, Q. Zhu, Y. Wang, X. Xu, and W. Yao, Nat. Phys. **13**, 356 (2016).
[11] M. Kögl, P. Soubelet, M. Brotons-Gisbert, A. V. Stier, B. D. Gerardot, and J. J. Finley, npj 2D Mater. Appl. **7**, 32 (2023).
[12] F. Miao, S.-J. Liang, and B. Cheng, npj Quantum Mater. **6**, 59 (2021).
[13] T. Peña, A. Dey, S. A. Chowdhury, A. Azizimanesh, W. Hou, A. Sewaket, C. Watson, H. Askari, and S. M. Wu, Appl. Phys. Lett. **122**, 143101 (2023).
[14] A. K. Katiyar and J. Ahn, Small Methods **9**, e2401404 (2025).
[15] Z. Bi, N. F. Q. Yuan, and L. Fu, Phys. Rev. B **100**, 035448 (2019).
[16] A. Imambekov, T. L. Schmidt, and L. I. Glazman, Rev. Mod. Phys. **84**, 1253 (2012).
[17] F. Escudero, A. Sinner, Z. Zhan, P. A. Pantaleón, and F. Guinea, Phys. Rev. Res. **6**, 023203 (2024).
[18] D. Zhai and W. Yao, Phys. Rev. Lett. **125**, 266404 (2020).
[19] D. Zhai, Z. Lin, and W. Yao, Rep. Prog. Phys. **87**, 108004 (2024).
[20] C. N. Lau, M. W. Bockrath, K. F. Mak, and F. Zhang, Nature **602**, 41 (2022).
[21] T. M. G. Mohiuddin *et al.*, Phys. Rev. B **79**, 205433 (2009).
[22] M. Kapfer *et al.*, Science **381**, 677 (2023).
[23] J.-B. Qiao, L.-J. Yin, and L. He, Phys. Rev. B **98**, 235402 (2018).
[24] F. Tagarelli *et al.*, Nat. Photonics **17**, 615 (2023).
[25] J. Kang, S. Tongay, J. Zhou, J. Li, and J. Wu, Appl. Phys. Lett. **102**, 012111 (2013).
[26] P. K. Nayak *et al.*, ACS Nano **11**, 4041 (2017).
[27] L. Zhang *et al.*, Nat. Commun. **11**, 5888 (2020).
[28] Y. Bai *et al.*, Nat. Mater. **19**, 1068 (2020).
[29] H. Zheng, D. Zhai, and W. Yao, 2D Mater. **8**, 044016 (2021).
[30] K. Tran *et al.*, Nature **567**, 71 (2019).
[31] W. Dieterich, I. Peschel, and W. R. Schneider, Z. Phys. B **27**, 177 (1977).
[32] W. Knorr, S. Brem, G. Meneghini, and E. Malic, Phys. Rev. Mater. **6**, 124002 (2022).
[33] H. Yu, G. B. Liu, J. Tang, X. Xu, and W. Yao, Sci. Adv. **3**, e1701696 (2017).
[34] Z. Sun, A. Ciarrocchi, F. Tagarelli, J. F. G. Marin, K. Watanabe, T. Taniguchi, and A. Kis, Nat. Photonics **16**, 79 (2022).